# A Big Data Lake for Multilevel Streaming Analytics

Ruoran Liu
School of Computing
Queen's University
Kingston, Canada
15rl1@queensu.ca

Haruna Isah
School of Computing
Queen's University
Kingston, Canada
h.isah@cs.queensu.ca

Farhana Zulkernine
School of Computing
Queen's University
Kingston, Canada
farhana@cs.queensu.ca

*Abstract*—Large organizations are seeking to create new architectures and scalable platforms to effectively handle data management challenges due to the explosive nature of data rarely seen in the past. These data management challenges are largely posed by the availability of streaming data at high velocity from various sources in multiple formats. The changes in data paradigm have led to the emergence of new data analytics and management architecture. This paper focuses on storing high volume, velocity and variety data in the raw formats in a data storage architecture called a data lake. First, we present our study on the limitations of traditional data warehouses in handling recent changes in data paradigms. We discuss and compare different open source and commercial platforms that can be used to develop a data lake. We then describe our end-to-end data lake design and implementation approach using the Hadoop Distributed File System (HDFS) on the Hadoop Data Platform (HDP). Finally, we present a real-world data lake development use case for data stream ingestion, staging, and multilevel streaming analytics which combines structured and unstructured data. This study can serve as a guide for individuals or organizations planning to implement a data lake solution for their use cases.

## I. INTRODUCTION

Big data is often described based on its five main aspects which include variety, velocity, volume, value and veracity. According to Mohamed et al. [1], variety refers to the different forms of big data types and sources such as structured, semi-structured and unstructured data generated on social media, Internet of Things (IoT), and smart-phones; velocity refers to the speed and frequency of the data generation; volume refers to the size of data being generated in gigabytes, terabytes and petabytes quantity; veracity refers to the extent data can be trusted, given the reliability of its source; and finally, value corresponds to variation in data values. Big data is playing a critical role in many application areas including healthcare, cybersecurity, business, and marketing. Big data analytics outcomes can help companies improve operations and uncover insights [2]. However, as data continues to flow from disparate sources, it is important to channel and maintain it in a central repository to enable effective data handling and analytics.

A data lake is an enterprise data management solution for storing vast amounts of data by bringing all the data into a single repository for further analysis without the restrictions of schema, security, or authorization [3]. It is a large storage repository that holds a vast amount of raw data in its native format until it is needed. This is in contrast to the traditional data warehouse approach, also known as schema on write, which requires more upfront design and assumptions about how the data will be used. In a data warehouse, adding data to a database requires data to be transformed into a pre-determined schema before it can be loaded. This step often consumes a lot of time, effort, and expense before the data can be used for downstream applications [4]. A data lake is meant to complement and not replace a data warehouse. At the core of a data lake is a set of repositories and components such as the traditional Relational Database Management System (RDBM) warehouses, operational data hubs, and distributed file system clusters [5].

There are several reasons why organizations are now considering the deployment of a data lake, which include the need to (i) define an infrastructure to manage repositories for all types of data i.e., structured, semi-structured or unstructured data with lineage and appropriate governance that will enable the organization to be compliant; (ii) provide a means for storing documents for specific analytical studies and reuse by multiple users including business users and data scientists; (iii) provide the necessary data lineage back to source systems, and (iv) enable users to access the analytical contents in a self-serve manner. For data science purposes, keeping all the data in a raw format is beneficial since often it is not clear upfront which data items may be valuable for a certain data analytics goal [5].

A data lake can either be designed from scratch or developed on top of existing platforms. An example of a data lake design from the scratch is Constance, a unified interface and query processing system by Hai et al. [4] for managing structural and semantic metadata, enriching metadata with schema, and matching and summarizing schema. An example of a data lake development on top of existing platforms is CUTLER, a system built on the Hadoop platform to gather data from a variety of sources for smart city analytics by Mehmood et al. [6]. Cloud platforms provide services that can be weaved together in an economical way to achieve the scalability requirement of a data lake. Examples of platforms that can be used to develop data lake include Amazon Web Services (AWS)[1], Microsoft Azure[2], Google Cloud [3], IBM Cloud [4], and Hadoop ecosystems (Cloudera, Hortonworks, and MapR) [7].

One of the main challenges of building a data lake is the understanding of the purpose of the system, analytics needs, and limitations of the possible components or technology that can be used to create the data lake [8]. Our industry-academic collaboration with a media monitoring and analytics company that serves clients from various sectors, focuses on developing an efficient and scalable data analytics and management framework, which can facilitate complex multilevel predictive analytics for real-time streaming data from a variety of sources including the Web and social media. This requires a data lake for the storage, analyses and query of high-velocity data.

---

[1] https://aws.amazon.com/
[2] https://azure.microsoft.com/
[3] https://cloud.google.com/
[4] https://www.ibm.com/cloud

We began with studying the existing technology options for developing the data lake. Following a thorough literature review, we implemented a data lake on top of the Hadoop Distributed File System (HDFS) to reliably store very large files across multiple clusters [9]. HDFS enables the use of the open-source Hadoop stack and does not enforce on adding a schema to the data flowing into the data lake [8].

The rest of this paper is organized as follows. Section II presents a background study on data warehouse, data lake, and platforms for data lake development. An overview of our Hadoop-based data lake is depicted in Section III. Section IV provides the data lake implementation details. The experimental and evaluation results are presented in Section V. Finally, Section VI presents a conclusion and a list of future work.

## II. BACKGROUND

The traditional data warehouse technology is based on the Extract, Transform and Load (ETL) data processing stages, which are no longer suitable for the extremely high rate at which data are emanating today from various sources [10]. The data lake has emerged as a tool that enables organizations to define, manage and govern the use of various big data technologies. The differences between a data warehouse and a data lake are outlined in TABLE 1[5].

TABLE I. DATA WAREHOUSE VS DATA LAKE

| Features | Data Warehouse | Data Lake |
|---|---|---|
| Data structure | Data is processed, only structured information is captured and organized in schemas | Data is raw, all data types (structured, semi-structured, unstructured) is captured in its original form |
| Users | Ideal for operational users such as business analyst since the data is structured and easy to use | Ideal for advanced users such as data scientists who carry out deep analysis with advanced analytical tools |
| Storage costs | Storing data is time-consuming and costly | Storing data is relatively inexpensive |
| Accessi-bility | Costly to make changes, thereby quite complicated | Updates can be made quickly thus making it highly accessible |
| Position of schema | Schema is defined before data is stored, thus offering performance and security | Schema is defined after data is stored, thus making it highly agile and scalable |
| Data processing | Uses the Extract, Transform, and Load (ETL) process. | Uses Extract, Load and Transform (ELT) process. |

---
[5] https://www.grazitti.com/blog/data-lake-vs-data-warehouse-which-one-should-you-go-for/

The flow and processing of data from various sources including social media and IoT are shown in Fig. 1 to illustrate the difference between a data warehouse and a data lake. The ETL process places the data in a schema as it stores (writes) the data to the relational database whereas the data lake stores the data in raw form. When any of the Hadoop applications use the data, the schema is applied to data as they are read from the data lake. The ETL step often discards some data as part of the process. In both the warehouse and Hadoop data lake cases, users access the data they need. However, in the case of Hadoop, it can happen as soon as the data are available in the data lake.

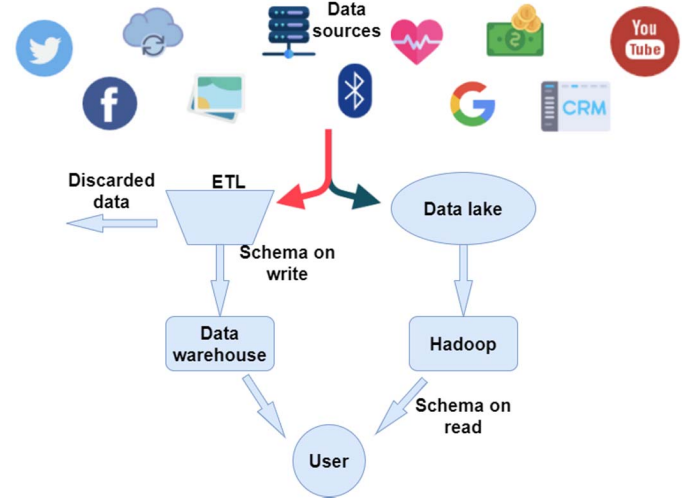

Fig. 1. Data warehouse versus data lake.

As mentioned in the introductory section, a data warehouse is a valuable business tool still in use in many data analytics pipelines. Data lake was developed to complement the data warehouse. The basic idea of a data lake is simple, all data collected by an organization will be stored in a single data repository in the lake in their original format. As such, the complex preprocessing and transformation of loading data into a warehouse is eliminated. Also, the upfront costs of data ingestion are reduced [11]. The major benefit of a data lake is the centralization of content from disparate sources. A data lake can have tens of thousands of tables or files and billions of records, which demands for scalable data storage, management and analytics. Once gathered at the data lake, data from multiple sources can be correlated, integrated and processed using leading edge big data analytics and search techniques which would have otherwise been impossible.

A data lake often contains proprietary or sensitive information that requires appropriate security measures. The security measures in a data lake can be implemented to allow partial access to selected information and to anonymized or encrypted data to users in a variety of roles, who do not have access to the original data. The lake should also allow anywhere anytime ubiquitous data access from around the globe. This feature will increase the re-use of the data and help organizations to more easily collect and process the data as required to drive business decisions. Allowing enterprise-wide role-based information access to a greater number of employees will also make an organization smarter, more agile, and innovative [12].

*A. Literature Survey*

According to Mendelevitch et al. [10], many existing pre-Hadoop data architectures tend to be rather strict and therefore, difficult to work with and make changes to. Fang et al. [3] described a data lake as a concept closely tied to Hadoop and its ecosystem of open source projects. According to Fang et al., data lake supports the following capabilities: (i) the capture and storage of large scale raw data at a low cost, (ii) the storage of many types of data in the same repository, (iii) ETL transformations and (iv) schema on read. A data lake is meant to address the daunting challenge of simplifying and facilitating the use of highly diverse big data to extract and provide knowledge.

Recently, Khine and Wang [11] described the primary concerns and challenges posed by a data lake. According to the authors, a data lake can provide value to various parts of an organization, but they are not "solution ready" for enterprise-wide data management. Besides, a data lake might have a tendency to data pollution, and the chances of it becoming a data swamp are high. This is because there is no veracity guarantee of all the data being fed into a data lake. As such, if no one knows what kind of data resides in the data lake, it might not be possible to find out whether some data in the lake are corrupted until it is too late. Therefore, preventing a data lake from becoming a data swamp is an interesting research topic. Next, we outline the strengths and weaknesses of previous studies aimed at developing a data lake and how our proposed system relates to or differs from these.

Hai et al. [4] proposed and evaluated a data lake that supports data ingestion from heterogeneous sources, allows metadata management over raw data, and query processing on both structured and semi-structured data. The system also embeds query rewriting engines and provides basic security and provenance mechanisms. However, the system needs to be extended to implement complex analytics functionality.

Walker and Alrehamy [13] proposed a unified storage facility for storing, analyzing and querying personal data and called it personal data lake. They describe how an individual can have a data lake that can collect and store all the digital data in a person's lifetime from emails, photos, medical records, invoices, bills, payments, certificates, and phone calls. They suggest the use of a graph database as an efficient way of storing metadata for the fast retrieval of matching data fragments. While this is an interesting work targeting data privacy and security, our goal in this study is to develop a data lake to support multilevel streaming analytics.

An interesting and recent study is the work by Mehmood et al. [6] which proposed and evaluated a data lake system on top of big data technologies to collect, store, and process heterogeneous data from various sources for smart city applications. Their proposed architecture is similar to our approach; however, their system is focused more on smart city data. Additionally, our architecture employs a variety of distributed technologies to support multilevel streaming analytics and to decouple the system components.

*B. Hadoop-based Data Lake*

One of the basic features of Hadoop is HDFS central storage space for inexpensive and redundant storage of large datasets at a much lower cost compared to the traditional storage systems. HDFS is at the core of the Hadoop data lake approach wherein all data are often stored in raw format, and what looks like the ETL step, is performed when the data is processed by Hadoop applications. No matter what kind of data needs processing, there is often a tool for importing data from or exporting data into HDFS. Once stored in the HDFS, data can be processed by any number of tools available in the Hadoop ecosystem. According to Mendelevitch et al. [10], the Hadoop-based data lakes offer the following three advantages over a more traditional approach.
1. All data are available, thereby eliminating the need to make any assumptions about future data use.
2. All data are sharable, as such multiple business units or researchers can use all available data including the ones that were probably previously not available due to data compartmentalization on disparate systems.
3. All-access methods are available, hence, processing engines such as MapReduce or Spark and query processing applications such as Hive or Spark SQL can be used to examine the data and process it as needed.

The advantages of Hadoop such as being open source, having support for various platforms and a large community base are what motivated us to choose Hadoop as a staging repository and HDFS as the main storage component of our data lake. Next, we compare three popular Hadoop-based platforms that are useful for developing a data lake. These platforms include the Hortonworks Data Platform (HDP), MapR, and Cloudera.

*C. Comparison of Hadoop-based Vendors*

Even though most of the software components that constitute the Hadoop ecosystem are open source, there are numerous benefits to using other vendors that offer Hadoop services beyond the openly accessible features either freely or with a commercial license. The number of vendors offering Hadoop distribution has thinned recently due to changes in the market [14]. Many vendors such as the IBM Cloud have resorted to reselling offerings from other vendors such as the HDP which has recently merged with or been acquired by Cloudera[6]. Currently, the three leading cloud platform providers of Hadoop-distribution are AWS, Microsoft, and Google. However, these platforms incur some fees and so our study is limited to the open-source offerings that can be used both on a private cloud or on-premise.

To determine the right Hadoop provider for our data lake, we embarked on comparisons of Hadoop distribution for the three major open-source Hadoop distribution vendors (HDP, MapR, and Cloudera) based on several key characteristics such as deployment models, enterprise-class features, security and data protection features, and support services that are important to have in a data lake. TABLE II describes these key features for the three different Hadoop distribution vendors.

---

[6] https://techcrunch.com/2019/01/03/cloudera-and-hortonworks-finalize-their-merger/

TABLE II. COMPARISON OF HADOOP-BASED VENDORS

| Features | Vendors | | |
|---|---|---|---|
| | *HDP* | *Cloudera* | *MapR* |
| Version | 3.1.5 | 6.3.3 | 6.1 |
| Deployment | Cloud-based, sandbox | Cloud-based, sandbox | Cloud-based, sandbox |
| Data storage | HDFS | HDFS, Kudu | MapR XD |
| Operational database | HBase | HBase | MapR Database |
| Search tools | HDP Search | Cloudera Search | MapR Search |
| Analytics tools | Spark, Hive, Pig, Storm, Drill, Kafka, Tez, Phonix, Impala, and Zeppelin | Cloudera Enterprise Data Hub, Essentials, Analytic DB, Operational DB and Data Science and Engineering | Spark, Drill, Hive, Tez, MapR Event Store, Impala |
| Security | Secured by Ranger | Secured by Sentry | Secured by default |

The most recent versions of the three vendors HDP, Cloudera, and MapR at the time of writing this paper are 3.1.5, 6.3.3, and 6.1 respectively. However, the current QuickStart VM version for Cloudera is 5.14 and it is not intended or supported for use in production. In terms of deployment, all three vendors offer cloud-based deployments and allow users to download distributions that can be deployed on-premise or in private clouds. The vendors also provide sandbox versions that can run in a virtual environment such as VMware, VirtualBox, Docker, and Kernel-based Virtual Machine (KVM) which is supported by Cloudera.

The main data storage in HDP is HDFS, a distributed Java-based file system, for storing large volumes of data. HDFS was designed to span large clusters of commodity servers and has demonstrated production scalability of up to 200 petabytes of storage and a single cluster of 4500 servers, supporting close to a billion files and blocks. Cloudera also uses HDFS but uses Kudu as storage for analytics on fast data. Kudu provides a combination of fast inserts and updates alongside efficient columnar scans to enable multiple real-time analytic workloads across a single storage layer. The main data storage in MapR is also a file system known as the MapR XD Distributed File and Object Store. MapR XD was designed to store data at an exabyte scale, support trillions of files, and combine analytics and operations into a single platform. Both HDP and Cloudera use HBase as a column-based NoSQL store for unstructured data whereas MapR has created an operational NoSQL database called MapR Database, as an alternative to HBase.

The search features in the three vendors HDP, Cloudera, and MapR are HDP Search, Cloudera Search, and MapR Search respectively. HDP Search uses Solr to offer a performant, scalable, and fault-tolerant enterprise search solution in an HDP cluster. HDP Search also uses connectors to index content from HDFS, HBase, Hive, Pig, Spark, and Storm in an HDP cluster. Cloudera Search incorporates Solr, which includes Lucene, SolrCloud, Tika, and Solr Cell. Cloudera Search provides easy, natural language access to data stored in or ingested into Hadoop, HBase, or cloud storage. MapR Search integrates LucidWorks search capabilities to perform full-text searches on data in a cluster without using a specialized query syntax.

HDP enables agile application deployment, machine learning and deep learning workloads, and real-time data warehousing using a variety of software tools such as Spark, Hive, Pig, Storm, and Phoenix. Cloudera offers software, services and supports both on-premise and across multiple cloud providers. The Cloudera products include (i) Cloudera Enterprise Data Hub, (ii) Cloudera Essentials, (iii) Cloudera Analytic DB, (iv) Cloudera Operational DB, and (v) Cloudera Data Science and Engineering. Finally, MapR supports a wide variety of analytics and data management tasks. It includes tools such as Spark, Drill, Hive on MR and Tez, MapR Event Store, Impala, JSON, and schema-less queries.

Security is essential for organizations that store and process sensitive data in the Hadoop ecosystem. Many organizations must adhere to strict corporate security policies. To ensure effective protection for the customers, HDP, Cloudera, and MapR use different approaches based on the core security features such as administration, authentication and perimeter security, authorization, audit, and data protection. HDP uses Ranger to provide centralized security administration and management. Cloudera uses a feature called Sentry to enforce authorization policies in a cluster. MapR unified platform is secured by default using built-in auditing, enterprise-grade encryption, expressive authorization, and flexible authentication. We opted to use HDP because of the ease of deployment and usability.

## III. PROPOSED DATA LAKE

The general architecture of the proposed Hadoop-based data lake is shown in Fig. 2. All data will be ingested into the data lake or staging repository such as HDFS in HDP. The ingestion can be batch, real-time or hybrid [2]. Many tools can be used to facilitate the ingestion of data from various sources into the data lake. Popular among the open-source options include NiFi[7], a data flow management system, and Kafka [8], a distributed messaging system for passing data between systems. These tools and their features have been outlined in [15]. Combining tools that do different things in better ways may allow for a build-up in functionality and increased flexibility in handling many use case scenarios. Sqoop is another tool designed for efficiently transferring bulk data between Hadoop and structured datastores such as relational databases.

The ingested content can then be analyzed using big data and analytics tools such as Spark. HBase and Hive are the commonly used tools in a Hadoop cluster for processing SQL queries. We used Hive as an ETL tool for batch inserts into HBase and to execute queries that join data in HBase tables with the data in HDFS files or external data stores. For high-performance analytics of both batch and streaming data, Spark is the best option. The analytics results can be visualized by various display tools such as Zeppelin via Web or mobile applications. The architecture also provides a possible solution to transport data from the existing database (relational database) to the HDFS and other Hadoop components to address the limitations of traditional computing.

---

[7] https://nifi.apache.org/

[8] https://kafka.apache.org/

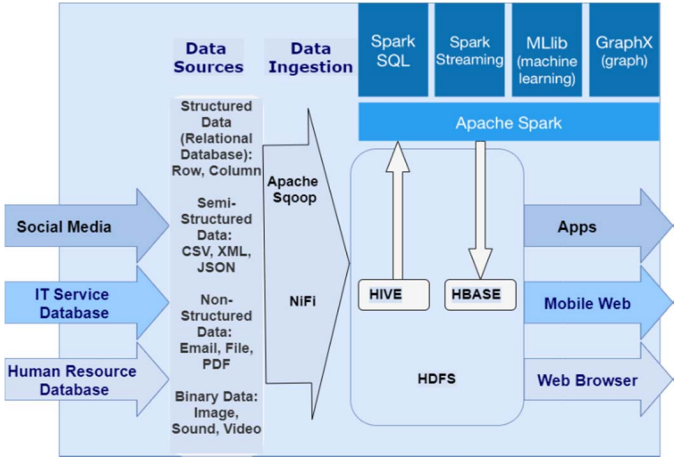

Fig. 2. General architecture and overview of the proposed data lake.

## IV. IMPLEMENTATION

### A. Use Case Scenario

We built our simple data lake architecture for the following use case scenario. A car trading company is striving to develop an efficient and scalable data lake for ingesting both structured and unstructured data from a variety of sources including relational databases, the Web and social media data. The goal was to provide a central data repository for the car inventory and related transactions as well as streaming data from the social media for business intelligence. This is an important use case for the future autonomous vehicle industry where data can be ingested from a variety of sensors in the cars, connected vehicles and devices. In this study we use simple data sources, which can be extended as needed to address more complex scenarios.

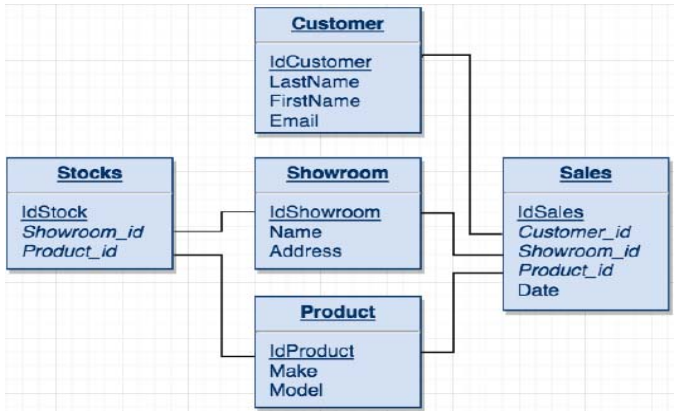

Fig. 3. The entity-relationship diagram of the car trading database [14].

### B. Data

The dataset for this implementation comprised an SQL database for the car trading firm [16] and Twitter data. The dataset was used to demonstrate a typical analytical use case for car sales and stocks. It was a MySQL database with a customer, product, showroom, sales and stock tables as demonstrated in Fig. 3. Streaming data from the Twitter API mentioning specific entities regarding the company and its car brands were used as the other source of data.

---

[9] https://cloud.soscip.org

### C. Implementation Environment

The project implementation environment was a Hadoop Hortonworks Data Platform (HDP) deployed on the SOSCIP Cloud[9], a cloud-accessible cluster of virtual machines having 48 GB RAM, 12 VCPU and 120 GB memory disk. Sqoop in HDP was used for transferring the relational data into the data lake while NiFi installed in the same cluster was used for ingesting streaming data using the Twitter Streaming API.

### D. Workflow

The enterprise data lake and big data architectures were built on Cloudera, which collected and processed all the raw data in one place, and then indexed that data into a Cloudera Search, Impala, and HBase for a unified search and analytics experience for end-users. Fig. 4 shows the implementation of the proposed Hadoop-based data lake. It supports (i) importing structured data from a relational database into HDFS using Sqoop, (ii) gathering unstructured data from Twitter into HDFS using NiFi, and (iii) using Spark SQL for data analysis, for example, to query the bestselling brands.

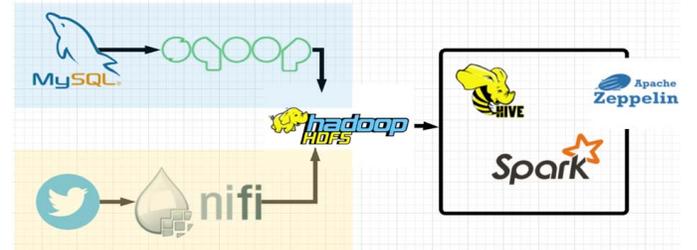

Fig. 4. Implementation of the Hadoop-based data lake.

We performed a simple multilevel analytics comprising of batch or static and streaming data ingestion using Sqoop and NiFi respectively at the first level followed by data staging in HDFS at the second level and finally, SQL and machine learning analytics in Hive and Spark as well as decision support and visualization at the highest levels. We started by loading the structured dataset into HDFS using Sqoop, followed by creating an analytical system in Hive to query the car trading data. Next, we used NiFi to connect to the Twitter Streaming API to gather tweets mentioning car brands specific to the company for sentiment analysis, a natural language processing technique that uses text analysis and computational linguistics to identify, extract, quantify, and study affective states and subjective information [17]. Apache Zeppelin was used as a multi-purposed web-based notebook for data exploration, visualization, sharing and collaboration in Hadoop and Spark.

## V. EXPERIMENTAL EVALUATIONS

To evaluate this work, we took some screenshots to describe some of the major steps of ingesting data into the data lake and the higher-level analytics that follow. For example, Fig. 5 shows the result of querying the bestselling brands from a table stored in Hive using SparkSQL, Fig. 6 shows tweet samples in JSON format, while Fig. 7 shows a comparison plot between the sales of the top ten brands and the frequency of each brand's name contained in the twitter messages.

```
brand1: String = Ford
brand2: String = Chevrolet
brand3: String = Dodge
brand4: String = Toyota
brand5: String = GMC
brand6: String = Mitsubishi
brand7: String = Mazda
brand8: String = Audi
brand9: String = Benz
brand10: String = Volkswagen
```

Fig. 5. Result of querying the bestselling brands.

```
{"tweet_id":"1109349236406140929","created_unixtime":1553324443328,"created_time":"Sat Mar 23 07:00:43
+0000 2019","lang":"en","location":"","displayname":"StunningCamer","time_zone":"","msg":"TechnaFit
Stainless 4 Brake Lines Blue for 200816 Mitsubishi EVO 10 LANCER https//tco/li30a9qfhI"}
{"tweet_id":"1109349239480561666","created_unixtime":1553324444061,"created_time":"Sat Mar 23 07:00:44
+0000 2019","lang":"en","location":"Brooklyn","displayname":"getraddielater","time_zone":"","msg":"RT
nytimes General Motors said that it would begin producing a new electric vehicle as part of its
Chevrolet lineup https//tco/Jqia2PMrN8"}
{"tweet_id":"1109349241112195082","created_unixtime":1553324444450,"created_time":"Sat Mar 23 07:00:44
+0000 2019","lang":"en","location":"","displayname":"StunningCamer","time_zone":"","msg":"TechnaFit
Stainless 4 Brake Lines Kit Clear for 200209 Audi A4 QUATTRO ALL https//tco/x4QiLSKq9F"}
```

Fig. 6. Tweet samples in JSON format.

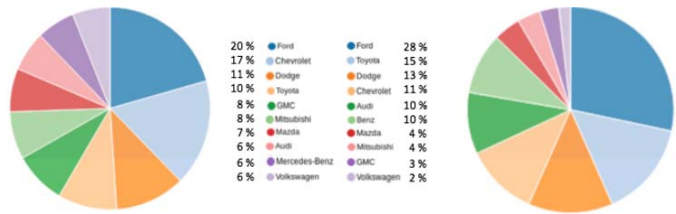

Fig. 7. Top ten brands and the frequency of mention in tweets.

The experiments illustrate the utility of implementing a data lake as a unified data management and analytics platform. The main idea of a data lake is to ingest raw data without processing and process data upon usage [2]. The proposed solution supports multilevel analytics which involves data ingestion, transformation, and storage of big data of various formats from a variety of sources. The use of HDP also provides additional enterprise features such as integrations to existing systems, robust security, data protection and governance.

## VI. CONCLUSIONS

The proliferation of the Internet and the evolution of social media and IoT have led to the problem of ingestion, processing, knowledge extraction, management and query processing involving big data. This study focuses on developing an end-to-end data lake solution for the ingestion, integration, and processing of both structured and unstructured data for multi-level analytics. The goal is to provide a platform for both data access and analytics to businesses and advanced data analytic users. First, we outlined the concepts regarding a big data lake, and the differences between a traditional data warehouse and a data lake. We described Hadoop as a data lake platform and compared three vendors offering Hadoop distribution services. The insights gained from the comparison enabled us to propose an end-to-end data lake using HDP. Finally, we developed and presented a data lake for a real-world use case for the automotive industry to demonstrate data stream ingestion, staging, and multilevel streaming analytics using both structured and unstructured data. This study can serve as a guide for individuals or organizations planning to implement a data lake solution for their use cases.

Future work will involve defining metadata when importing relational data into HDFS using Sqoop. It will also involve storing the original data into different NoSQL databases to compare the computation time.

ACKNOWLEDGMENT

We would like to express a special thanks to the Southern Ontario Smart Computing for Innovation Platform (SOSCIP), Ontario Centres of Excellence (OCE) and IBM Canada for supporting this research.